\documentclass[smallextended]{svjour3}     % onecolumn (second format)
\smartqed  % flush right qed marks, e.g. at end of proof
\usepackage{graphicx}
\usepackage{amsmath}
\usepackage{amssymb}
\usepackage{esint}%
\usepackage{multirow}
\begin{document}

\title{Testing the interaction of dark energy to dark matter through the
analysis of virial relaxation of clusters Abell Clusters A586 and
A1689 using realistic density profiles%Grants or other notes
}

\titlerunning{Testing DE-DM interaction with A586 and A1689}        % if too long for running head

\author{Orfeu Bertolami   \and
	Francisco. Gil Pedro   \and
        Morgan Le Delliou %etc.
}

\authorrunning{O. Bertolami,F. Gil Pedro \& M. Le Delliou} % if too long for running head

\institute{O. Bertolami \at
Departamento de F\'{\i}sica e Astronomia, Faculdade de Ci\^encias
da Universidade do Porto\\
Rua do Campo Alegre 687, 4169-007 Porto, Portugal\\
%              first address \\
%              Tel.: +123-45-678910\\
%              Fax: +123-45-678910\\
              \email{orfeu.bertolami@fc.up.pt  fgpedro@fisica.ist.utl.pt%fauthor@example.com
}             \\
             \emph{O.B. Also at:} Instituto de Plasmas e Fus\~ao Nuclear,
Instituto Superior T\'ecnico, Avenida Rovisco Pais 1, 1049-001 Lisboa,
Portugal%of F. Author  %  if needed
           \and
           F. Gil Pedro \at
Rudolf Peierls Centre for Theoretical Physics\\
1 Keble Road Oxford OX1 3NP, UK\\
              \email{f.pedro1@physics.ox.ac.uk%fauthor@example.com
           \and
           M. Le Delliou \at
              Instituto de Física Teórica UAM/CSIC, 
Facultad de Ciencias, C-XI,
Universidad Aut\'onoma de Madrid,
 Cantoblanco, 28049 Madrid SPAIN
              \email{Morgan.LeDelliou@uam.es}             \\
             \emph{Also at:} Centro de de Astronomia e Astrof\'{\i}sica da Universidade de Lisboa, Faculdade de Ci\^encias da Universidade de Lisboa
Edif\'{\i}cio C8, Campo Grande
P-1749-016 LISBOA
Portugal
}
}
\date{Received: date / Accepted: date}

\maketitle

\begin{abstract}
Interaction between dark energy and dark matter is probed through
deviation from the virial equilibrium for two relaxed clusters: A586
and A1689. The evaluation of the virial equilibrium is performed using
realistic density profiles. The virial ratios found for the more realistic
density profiles are consistent with the absence of interaction.

 \PACS{98.80.-k \and 98.80.Cq %\hfill \emph{Preprint:} DF/IST-4.2007%\and more
}
\end{abstract}

\section{Introduction}
\label{intro}

Searching for evidence and possible strategies for detection of dark
energy and dark matter are among the most pressing issues of contemporary
physics. Dark energy (DE) and dark matter (DM) are the fundamental
building blocks of the cosmological standard model based on general
relativity. If on one hand, the interaction of the dark sector with
the standard model such as neutrinos \cite{Bernardini:2008pn} and
the Higgs field \cite{Silveira:1985rk} lead to well defined experimental
implications, such as for instance, a possible deviation from the
Rutherford-Soddy radiative decay law \cite{Bento:2009im} (see eg.
Ref. \cite{Bento:2009xa} for general discussions), on another hand,
interaction within the dark sector itself should not be excluded.
Indeed, a recent study \cite{Barreiro:2010nb} suggests that a putative
interaction might also be detectable using for instance gamma-ray
bursts. The interaction between dark energy and dark matter can be
introduced either by ad hoc arguments \cite{Amendola:1999er} or because
dark energy and dark matter are unified in the context of some framework
\cite{Kamenshchik:2001cp,Bento:2002ps,Bertolami:2007wb}. Of course
it is quite relevant to search for observational evidence of this
interaction.

In a previous work \cite{Bertolami:2007zm} we have analysed the effect
of the interaction between dark energy and dark matter on the virial
equilibrium of clusters. Our method is based on the generalisation
of the Layzer-Irvine equation, the cosmic virial theorem equation,
which was then applied to the Abell cluster A586. This cluster was
chosen given that it is presumably in equilibrium and has not undergone
interactions for quite a long time \cite{Cypriano:2005}. We have
also argued that, based on the analysis of the bias parameter, that
the dark sector interactions do signal a possible violation of the
Equivalence Principle at cosmological scale \cite{Bertolami:2007zm,Bertolami:2007tq}.
This proposal has been extended for other clusters in Refs. \cite{Abdalla},
and as in our A586 analysis, evidence for DE-DM interaction was encountered.

In this work we reexamine the cluster A586 and extend our analysis
to the Abell cluster A1689 considering that mass distribution within
the cluster is not constant. The additional cluster A1689 is chosen
as it shares with A586 the feature of showing a negligible amount
of mergers, as inferred from its low X-ray substructure \cite{Lagana}.
We consider four distinct density profiles: Navarro, Frenk and White
(NFW) profile \cite{Navarro:1995iw}, isothermal sphere (ISO) profile
\cite{Larson69,Penston69,GunnGott72,FG84,Bert85,Peebles80}, Moore's
(M) profile \cite{Moore99} and Einasto profile \cite{Einasto65,Einasto68,Einasto69}.
We also obtain results for up to four different techniques of evaluating
the velocity dispersion $\sigma_{v}$ of the haloes. We argue that
the most relevant technique for determination of the velocity dispersion
is through the X-ray temperature. Our analysis reveals that the use
of more realistic mass profiles brings the virial ratio close to its
canonical value in the absence of interaction.

This work is organized as follows: in section \ref{sec:Computations-of-the}
we describe the techniques used for fitting the data with the various
density profiles; section \ref{sec:Analysis-of-A586} presents our
analysis of the virial ratios for the previously used cluster A586;
section \ref{sec:Analysis-of-the} considers the A1689 cluster; in
section \ref{sec:Interaction-and-violation} we interpret our results
in terms of DE-DM interaction. Finally, in section \ref{sec:Conclusions}
we present our conclusions.\\
\\
\\
\\
\\

\section{\label{sec:Computations-of-the}Computations of the Virial ratio
for a given density profile}

In this section we present our method and summarise the results of
applying realistic density profiles to estimate the interaction between
DE and DM in relaxed systems as discussed in Refs. \cite{Bertolami:2007zm,Bertolami:2007tq}.
We will focus on two relaxed galaxy clusters: A586, which was already
analysed in Ref. \cite{Bertolami:2007zm} with a top-hap density profile,
and A1689.

We start by defining the basic quantities and deriving the necessary
relations between them. In particular we focus on obtaining various
density profile fits for the haloes, and using the photometric coordinates
of the galaxies, we compute the average distance between a galaxy
and the centre of the cluster, $\langle R\rangle$, at which point
we are in condition to evaluate the various density profiles parameters
and the corresponding errors for the cluster A586. For the cluster
A1689, we use weak lensing surface density data. To compute the virial
ratio $\rho_{K}/\rho_{W}$ we must also find the velocity dispersion
$\sigma_{v}$. From Refs. \cite{Cypriano:2005,Lemze:2007gh,Hoekstra:2007nc,Girardi:2000zc}
we obtain the weak lensing and photometric data necessary to compute
the different density profiles as well as values for $\sigma_{v}$,
as given by different methods; note that we approximate the velocity
dispersion by its average value therefore considering it to be constant
over the cluster ; we can then compute the ratio of kinetic energy
to potential energy using different data sources (see Table \ref{tab:Velocity-dispersions}
for A586 and Table \ref{tab:Summary-of-A1689} for A1689). 

In Ref. \cite{Bertolami:2007zm} we have used the simplest approach
of a Top-Hat density profile and used the velocity dispersion estimated
from weak lensing measurements. For the purpose of comparison, we
present in Table \ref{tab:Virial-ratio-Master} the outputs of such
profile with the same velocity dispersion as for the other more realistic
profiles. Although that analysis allows for claiming  detection \cite{Bertolami:2007zm},
a more accurate estimate of the density leads to an underestimation
of the potential energy by placing more mass far from the centre than
a realistic decreasing profile. This implies an overestimate of the
magnitude of the virial ratio, as can be seen for central values in
Table \ref{tab:Virial-ratio-Master}. This should be corrected with
the use of more realistic density profiles.

It is important to stress that the methods used vary slightly for
the two clusters we are studying here. For the A586 we proceed in
the same spirit as in Ref. \cite{Bertolami:2007zm}, estimating the
density and shape parameters for the several density profiles from
total 3D mass and galaxy coordinates. On the other hand, for the A1689
cluster we fit directly the profiles to the 2D surface density obtained
from lensing measurements \cite{Lemze:2007gh}.

\subsection{NFW profile}

Given its prominence in discussions about the density profile emerging
from realistic N-body simulations in the context of the $\Lambda$CDM
model, the NFW density profile \cite{Navarro:1995iw} will be used.
It reads 
\begin{equation}
\rho(r)=\frac{\rho_{0}}{\frac{r}{r_{0}}\left(1+\frac{r}{r_{0}}\right)^{2}},\label{eq:NFW}
\end{equation}
 where $\rho_{0}$ and $r_{0}$ are the density and shape parameters
respectively. Note that since $\rho$ only depends on $r$, spherical
symmetry is built into these computations from the very start.

In the analysis of the cluster A1689, the 2D surface density from
lensing measurements is used to perform a fit in order to obtain the
estimates for $\rho_{0}$ and $r_{0}$. For the cluster A586 we proceed
by computing the mass by integrating Eq. (\ref{eq:NFW}) over the
volume: 
\begin{equation}
M  \equiv4\pi\int_{0}^{R}\frac{\rho_{0}r^{2}}{\frac{r}{r_{0}}\left(1+\frac{r}{r_{0}}\right)^{2}}dr
 =4\pi r_{0}^{3}\rho_{0}\left[\ln\left(1+\frac{R}{r_{0}}\right)-\frac{R}{R+r_{0}}\right].
\label{eq:MNFW}
\end{equation}

The mean radius $\langle R\rangle$ can then be defined with 
\begin{equation}
M\left\langle R\right\rangle   \equiv4\pi\int_{0}^{R}\frac{\rho_{0}r^{3}}{\frac{r}{r_{0}}\left(1+\frac{r}{r_{0}}\right)^{2}}dr
  =4\pi r_{0}^{4}\rho_{0}\left[\frac{R}{r_{0}}-2\ln\left(1+\frac{R}{r_{0}}\right)+\frac{R}{R+r_{0}}\right],
\label{eq:meanR}
\end{equation}
 which after considering Eq. (\ref{eq:MNFW}) can be written as follows:

\begin{equation}
\langle R\rangle=r_{0}\frac{\left[\frac{R}{r_{0}}-2\ln\left(1+\frac{R}{r_{0}}\right)+\frac{R}{R+r_{0}}\right]}{\left[\ln\left(1+\frac{R}{r_{0}}\right)-\frac{R}{R+r_{0}}\right]}.\label{eq:meanR2}
\end{equation}

We can then get the shape parameter $r_{0}$ for A586 by numerically
inverting $\left\langle R\right\rangle $. We then use it to define
the NFW density parameter $\rho_{0}$ from the observed mass $M$
contained within the radius $R$ by inverting Eq. (\ref{eq:MNFW}):
\begin{equation}
\rho_{0}=\frac{M}{4\pi r_{0}^{3}\left[\ln\left(1+\frac{R}{r_{0}}\right)-\frac{\frac{R}{r_{0}}}{1+\frac{R}{r_{0}}}\right]}.\label{eq:rho0}
\end{equation}

The kinetic energy density is defined in terms of the average velocity
dispersion, $\sigma_{v}$, to be 
\begin{equation}
\rho_{K}  =\frac{1}{2}\frac{3}{V}\int\rho\sigma_{v}^{2}dV
  =\frac{9}{2}\frac{r_{0}^{3}}{R^{3}}\rho_{0}\int_{0}^{R}\frac{r}{\left[r+r_{0}\right]^{2}}\sigma_{v}^{2}dr.
\end{equation}
 Assuming a constant average velocity dispersion, it reads 
\begin{equation}
\rho_{K}  =\frac{1}{2}\frac{M}{V}3\sigma_{v}^{2}
  =\frac{9}{2}\frac{r_{0}^{3}}{R^{3}}\rho_{0}\left[\ln\left(1+\frac{R}{r_{0}}\right)-\frac{R}{R+r_{0}}\right]\sigma_{v}^{2}
  =\frac{9}{8\pi}\frac{M}{R^{3}}\sigma_{v}^{2},
\label{eq:rhoK}
\end{equation}
 as used in Ref. \cite{Bertolami:2007zm}.

The potential energy is given by 
\begin{align}
\rho_{W} & =-\frac{4\pi}{4\pi R^{3}/3}\int_{0}^{R}\frac{\rho(r)GM(r)}{r}r^{2}dr\\
& =-\frac{3GM^{2}}{4\pi R^{3}r_{0}}\frac{\left[\left(1+\frac{R}{r_{0}}\right)\left\{ \frac{1}{2}\left(1+\frac{R}{r_{0}}\right)-\ln\left(1+\frac{R}{r_{0}}\right)\right\} -\frac{1}{2}\right]}{\left[\left(1+\frac{R}{r_{0}}\right)\ln\left(1+\frac{R}{r_{0}}\right)-\frac{R}{r_{0}}\right]^{2}}.\label{eq:rhoW}
\end{align}

As discussed in Ref. \cite{Bertolami:2007zm}, the existence of interaction
between DE and DM is estimated by comparing the ratio $\rho_{K}/\rho_{W}$
with the expected $-1/2$ value arising from the virial theorem. Taking
the ratio of Eqs. (\ref{eq:rhoK}) and (\ref{eq:rhoW}) we find: 
\begin{equation}
\frac{\rho_{K}}{\rho_{W}}=-\frac{\frac{3}{2}\frac{r_{0}}{GM}\sigma_{v}^{2}\left[\left(1+\frac{R}{r_{0}}\right)\ln\left(1+\frac{R}{r_{0}}\right)-\frac{R}{r_{0}}\right]^{2}}{\left[\left(1+\frac{R}{r_{0}}\right)\left\{ \frac{1}{2}\left(1+\frac{R}{r_{0}}\right)-\ln\left(1+\frac{R}{r_{0}}\right)\right\} -\frac{1}{2}\right]}.\label{eq:ratio}
\end{equation}
Throughout our analysis, the errors of the ratios are computed according
to the following equation: 
\begin{align}
\Delta\left(\frac{\rho_{K}}{\rho_{W}}(\left\{ o_{i}\right\} )\right) & =\sqrt{\sum_{p_{i}\in\left\{ o_{i}\right\} }\left(\frac{\partial\rho_{K}/\rho_{W}}{\partial p_{i}}\Delta(p_{i})\right)^{2}},\label{eq:errRatio}
\end{align}
where $\left\{ o_{i}\right\} $ are the parameters carrying measurement
errors and are defined for each cluster as $\left\{ o_{i}\right\} =\left\{ \sigma_{v},M\right\} $
for A586 and as $\left\{ o_{i}\right\} =\left\{ \sigma_{v},r_{0},\rho_{0}\right\} $
for A1689.

\subsection{Isothermal profile}

Introduced first as the natural outcome of spherically symmetric DM
self-gravitating infall \cite{Larson69,Penston69,GunnGott72,FG84,Bert85,Peebles80},
the isothermal  density profile is simpler than the NFW one: 
\begin{align}
\rho= & \frac{\rho_{0}}{\left(\frac{r}{r_{0}}\right)^{2}}.\label{eq:Iden}
\end{align}
Here the fiducial radius $r_{0}$ and density $\rho_{0}$, or mass
$M_{0}=4\pi/3\rho_{0}r_{0}^{3}$, are arbitrary as the profile is
self-similar: there is no characteristic scale, so we can chose the
mass and total radius of the halo as the fiducial values,%\foreignlanguage{british}{
\begin{align}
M_{0} & =M,\\
r_{0} & =R.
\end{align}
%}

The mass is found by integrating Eq. (\ref{eq:Iden}) over the volume:
\begin{equation}
M  \equiv4\pi\int_{0}^{R}\frac{\rho_{0}r^{2}}{\left(\frac{r}{r_{0}}\right)^{2}}dr
  =4\pi r_{0}^{2}R\rho_{0}=M_{0}\frac{R}{r_{0}}.
\label{eq:MI}
\end{equation}

The mean radius $\langle R\rangle$ can then be defined with 
\begin{equation}
M\left\langle R\right\rangle   \equiv4\pi\int_{0}^{R}\frac{\rho_{0}r^{3}}{\left(\frac{r}{r_{0}}\right)^{2}}dr
  =2\pi r_{0}^{2}\rho_{0}R^{2},
\label{eq:ImeanR}
\end{equation}
 which after considering Eq. (\ref{eq:MI}) can be written as follows:

\begin{equation}
\langle R\rangle=\frac{R}{2}.\label{eq:ImeanR2}
\end{equation}

The new kinetic energy density is now defined to be 
\begin{align}
\rho_{K} & =\frac{1}{2}\frac{3}{V}\int\rho\sigma_{v}^{2}dV=\frac{9}{2}\frac{r_{0}^{2}}{R^{3}}\rho_{0}\int_{0}^{R}\sigma_{v}^{2}dr%\nonumber \\
 & =\frac{9}{8\pi}\frac{M_{0}}{R^{3}r_{0}}\int_{0}^{R}\sigma_{v}^{2}dr.
\end{align}
 Assuming a constant average velocity dispersion, it reads 
\begin{equation}
\begin{split}\rho_{K} & =\frac{1}{2}\frac{M}{V}3\sigma_{v}^{2}=\frac{9}{8\pi}\frac{M_{0}}{R^{3}r_{0}}R\sigma_{v}^{2}=\frac{9}{8\pi}\frac{M}{R^{3}}\sigma_{v}^{2},\end{split}
\label{eq:IrhoK}
\end{equation}
again as in Ref. \cite{Bertolami:2007zm}.

The potential energy is given by%\foreignlanguage{british}{
\begin{align}
\rho_{W} & =-\frac{4\pi}{4\pi R^{3}/3}\int_{0}^{R}\frac{\rho(r)GM(r)}{r}r^{2}dr%\nonumber \\
 & =-\frac{3GM_{0}^{2}}{4\pi R^{2}r_{0}^{2}}=-\frac{3GM^{2}}{4\pi R^{4}}.\label{eq:IrhoW}
\end{align}
%} 

Taking the ratio of Eqs. (\ref{eq:IrhoK}) and (\ref{eq:IrhoW}) we
find: 
\begin{equation}
\frac{\rho_{K}}{\rho_{W}}=-\frac{3}{2}\frac{r_{0}}{GM_{0}}\sigma_{v}^{2}=-\frac{3}{2}\frac{R}{GM}\sigma_{v}^{2}~.\label{eq:Iratio}
\end{equation}

\subsection{Moore's profile}

We consider now the Moore density profile that also arises as a suitable
dark halo profile from N-body simulations \cite{Moore99}. It is believed
to be quite accurate to describe galaxy size halo formation \cite{JingSuto}:
%\foreignlanguage{british}{
\begin{align}
\rho= & \frac{\rho_{0}}{\left(\frac{r}{r_{0}}\right)^{\frac{3}{2}}\left[1+\left(\frac{r}{r_{0}}\right)^{\frac{3}{2}}\right]},\label{eq:Mden}
\end{align}
%}
where $\rho_{0}$ and $r_{0}$ are the new mass density and shape
parameters, respectively. The mass is found by integrating Eq. (\ref{eq:Mden})
over the volume:%\foreignlanguage{british}{
\begin{align}
M & =4\pi\rho_{0}\int_{0}^{R}\frac{r^{2}}{\left(\frac{r}{r_{0}}\right)^{\frac{3}{2}}\left[1+\left(\frac{r}{r_{0}}\right)^{\frac{3}{2}}\right]}dr%\nonumber \\
 & =\frac{8\pi}{3}r_{0}^{3}\rho_{0}\ln\left(1+\left(\frac{R}{r_{0}}\right)^{\frac{3}{2}}\right).\label{eq:MM}
\end{align}
%}
The mean radius $\langle R\rangle$ can then be defined with 
\begin{eqnarray}
M\left\langle R\right\rangle  & = & 4\pi r_{0}^{3}\rho_{0}\int_{0}^{R}\frac{\left(\frac{r}{r_{0}}\right)^{\frac{3}{2}}}{1+\left(\frac{r}{r_{0}}\right)^{\frac{3}{2}}}dr,\label{eq:MmeanR}
\end{eqnarray}
and considering Eq. (\ref{eq:MM}), it can be written as:

%\selectlanguage{british}%
\begin{multline}
\left\langle R\right\rangle =\frac{3}{2}r_{0}\left[\frac{R}{r_{0}}+\frac{\pi}{3\sqrt{3}}+\frac{1}{3}\ln\left(\frac{1-\sqrt{\frac{R}{r_{0}}}+\frac{R}{r_{0}}}{\left(1+\sqrt{\frac{R}{r_{0}}}\right)^{2}}\right)\right.\\
\left.\vphantom{\left(\frac{1-\sqrt{\frac{R}{r_{0}}}+\frac{R}{r_{0}}}{\left(1+\sqrt{\frac{R}{r_{0}}}\right)^{2}}\right)}-\frac{2}{\sqrt{3}}\arctan\left[\frac{2\sqrt{\frac{R}{r_{0}}}-1}{\sqrt{3}}\right]\right]\textrm{\Huge{\bf /}}\ln\left(1+\left(\frac{R}{r_{0}}\right)^{\frac{3}{2}}\right).\label{eq:MmeanR2}
\end{multline}

%\selectlanguage{english}%
We can get the shape parameter $r_{0}$ numerically by inverting $\left\langle R\right\rangle $,
which can be then used to define the Moore density parameter, $\rho_{0}$,
with the observed mass, $M$, contained within the radius, $R$, by
inverting Eq. (\ref{eq:MM}):%\foreignlanguage{british}{
\begin{align}
\rho_{0} & =\frac{3M}{8\pi r_{0}^{3}\ln\left(1+\left(\frac{R}{r_{0}}\right)^{\frac{3}{2}}\right)}.\label{eq:Mrho0}
\end{align}
%}

Thus, the kinetic energy density is defined to be %\foreignlanguage{british}{
\begin{align}
\rho_{K} & =\frac{1}{2}\frac{3}{V}\int\rho\sigma_{v}^{2}dV%\nonumber \\
 & =\frac{9}{2}\frac{\rho_{0}r_{0}^{3}}{R^{3}}\int_{0}^{R}\frac{r^{\frac{1}{2}}}{r^{\frac{3}{2}}+r_{0}^{\frac{3}{2}}}\sigma_{v}^{2}dr.
\end{align}
%} 
 Assuming a constant average velocity dispersion, it reads %\foreignlanguage{british}{
\begin{align}
\rho_{K} & =\frac{1}{2}\frac{M}{V}3\sigma_{v}^{2}=3\rho_{0}\left(\frac{r_{0}}{R}\right)^{3}\ln\left(1+\left(\frac{R}{r_{0}}\right)^{\frac{3}{2}}\right)\sigma_{v}^{2}%\nonumber \\
 & =\frac{9}{8\pi}\frac{M}{R^{3}}\sigma_{v}^{2},\label{eq:MrhoK}
\end{align}
%}
still as in Ref. \cite{Bertolami:2007zm}.

The potential energy is given by %\foreignlanguage{british}{
\begin{align}
\rho_{W} & =-\frac{4\pi}{4\pi R^{3}/3}\int_{0}^{R}\frac{\rho(r)GM(r)}{r}r^{2}dr%\nonumber \\
 & =-\frac{8\pi Gr_{0}^{5}\rho_{0}^{2}}{R^{3}}\int_{0}^{\frac{R}{r_{0}}}\frac{\ln\left(1+X^{\frac{3}{2}}\right)}{X^{\frac{1}{2}}\left(1+X^{\frac{3}{2}}\right)}dX\nonumber \\
 & =-\frac{8\pi Gr_{0}^{5}\rho_{0}^{2}}{R^{3}}F\left(\frac{R}{r_{0}}\right).\label{eq:MrhoW}
\end{align}
%}
Taking the ratio of Eqs. (\ref{eq:MrhoK}) and (\ref{eq:MrhoW})
we find: 
\begin{gather}
\frac{\rho_{K}}{\rho_{W}}=-\frac{r_{0}}{GM}\frac{\ln^{2}\left(1+\left(\frac{R}{r_{0}}\right)^{\frac{3}{2}}\right)}{F\left(\frac{R}{r_{0}}\right)}\sigma_{v}^{2}\label{eq:Mratio}
\end{gather}

\subsection{Einasto's profile}

The Einasto density profile \cite{Einasto65,Einasto68,Einasto69}
was originally used to describe the internal density profiles of galaxies
and has been proposed as a better fit model for $\Lambda$CDM haloes
\cite{Graham:2005xx,Graham:2006ae,Lapi:2010is}: 
\begin{align}
\rho & =\rho_{e}\exp\left[-d_{n}\left(\left(\frac{r}{r_{e}}\right)^{\frac{1}{n}}-1\right)\right]%\\
 & =\rho_{0}\exp\left[-2n\left(\frac{r}{r_{-2}}\right)^{\frac{1}{n}}\right],\label{eq:EinDen}
\end{align}
where $\rho_{0}=\rho_{e}e^{d_{n}}$ is the central density and $r_{-2}$
the radius at which the slope $d\ln\rho/d\ln r=-2$, the isothermal
value. The radius $r_{e}$ is defined such that it contains half the
total mass and $d_{n}$ is an integration boundary to ensure that.
$n$ gives the strength of the density fall. From Eq. (\ref{eq:EinDen})
the corresponding mass then reads:
\begin{align}
M & =4\pi\int_{0}^{R}\rho_{0}\exp\left[-2n\left(\frac{r}{r_{-2}}\right)^{\frac{1}{n}}\right]r^{2}dr.\label{eq:EinM}
\end{align}
The mean radius $\langle R\rangle$ then becomes defined with
\begin{align}
M\left\langle R\right\rangle & =4\pi\int_{0}^{R}\rho_{0}\exp\left[-2n\left(\frac{r}{r_{-2}}\right)^{\frac{1}{n}}\right]r^{3}dr.\label{eq:RmeanEinasto}
\end{align}
We can get the shape parameter $r_{-2}$ from observing the mean intergalactic
distance and numerically solving Eq. (\ref{eq:RmeanEinasto}) together
with Eq. (\ref{eq:EinM}).

We can then use it to compute the central density parameter $\rho_{0}$
by making use of Eq. (\ref{eq:EinM}). The kinetic and potential energy
densities are computed from their definition
\begin{equation}
\begin{aligned}\rho_{K} & =\frac{1}{2}\frac{3}{V}\int\rho\sigma_{v}^{2}dV\rho_{W},%\\
&\rho_{W} & =-\frac{4\pi}{4\pi R^{3}/3}\int_{0}^{R}\frac{\rho(r)GM(r)}{r}r^{2}dr,
\end{aligned}
\end{equation}
through numerical integration. Note that it is also possible to perform
the integrations analytically as done for the previous cases and find
expressions relating the profile parameters and the virial ratio to
the data $M,\langle R\rangle$ and $\sigma$. However, since these
expressions for the Einasto profile are not particularly illuminating
we will omit them.

\begin{table}
\caption{\label{tab:Velocity-dispersions}Velocity dispersions from the various
observations of A586 as given by Ref. \cite{Cypriano:2005}.}%\begin{centering}
\begin{tabular}{ll}
\hline 
Method  & $\sigma$ (Km/s) \tabularnewline
\hline 
X-ray Luminosity  & $1015\pm500$ \tabularnewline
X-ray Temperature  & $1174\pm130$ \tabularnewline
Weak lensing  & $1243\pm58$ \tabularnewline
Velocity distribution  & $1161\pm196$ \tabularnewline
\hline 
\end{tabular}
%\par\end{centering}

\end{table}

\section{\label{sec:Analysis-of-A586}Analysis of A586 cluster}

\subsection{Data Analysis}

%\subsection{Top-hat profile}

In the analysis of A586 we used the same data used in the original
analysis performed in \cite{Bertolami:2007zm}, that is: 
\begin{itemize}
\item member galaxy coordinates 
\item total mass inside a radius R, obtained from weak lensing measurementes
\cite{Cypriano:2005}: 
\begin{align}
M & =(4.3\pm0.7)\times10^{14}M_{\odot} &
R & =422\: Kpc
\label{eq:}
\end{align}
 
\item velocity dispersion from different sources, table \ref{tab:Velocity-dispersions}. 
\end{itemize}
Using the coordinates of the galaxies that compose A586 we can compute
the average distance between a galaxy and the centre of the cluster,
$\langle R\rangle$. We start by determining the centre of the cluster.
This is done by computing the average declination and right ascension
from the coordinates of the 31 galaxies in the cluster.

The distance of a galaxy i, with coordinates $(\alpha_{i},\delta_{i})$,
to the centre of the cluster $(\alpha_{c},\delta_{c})$ is given by:
\begin{equation}
r_{i}^{2}=2d^{2}[1-cos(\alpha_{i}-\alpha_{c})cos(\delta_{c})cos(\delta_{i})-sin(\delta_{c})sin(\delta_{i})].\label{eq:r2}
\end{equation}
Using Eq. (\ref{eq:r2}) to compute $r_{i}$ for all galaxies in the
reduced sample and then taking the average: 
\begin{equation}
\langle R\rangle=223.6\: Kpc\,.\label{eq:R}
\end{equation}

We can then proceed to compute the shape parameters of the various
density profiles as described in the previous section and through
them compute the virial ratio and the interaction between DE and DM.

\begin{table}
\caption{\label{tab:Virial-ratio-Master} Virial ratio from the various observations
of A586 obtained from data of Ref. \cite{Cypriano:2005} using different
density profiles.}
\begin{tabular}{l|l|l|l|l}
\hline 
Method  & Top Hat  & NFW  & Isothermal  & \tabularnewline
\hline 
X-ray Luminosity  & $-0.516\pm0.516$  & $-0.408\pm0.407$  & $-0.353\pm0.352$  & \tabularnewline
X-ray Temperature  & $-0.691\pm0.190$ & $-0.545\pm0.150$  & $-0.472\pm0.130$  & \tabularnewline
Weak lensing  & $-0.774\pm0.145$ & $-0.611\pm0.115$  & $-0.529\pm0.099$  & \tabularnewline
Velocity distribution  & $-0.676\pm0.253$  & $-0.533\pm0.200$  & $-0.461\pm0.173$  & \tabularnewline
\hline 
\end{tabular}

%\centering{}%
\begin{tabular}{l|l|l|l}
\hline 
 & Moore  & Einasto $n=1$  & Einasto $n=6$ \tabularnewline
\hline 
 & $-0.316\pm0.315$  & $-0.416\pm0.415$ & $-0.405\pm0.404$\tabularnewline
 & $-0.423\pm0.116$ & $-0.556\pm0.153$ & $-0.542\pm0.149$\tabularnewline
 & $-0.474\pm0.089$  & $-0.624\pm0.117$ & $-0.608\pm0.114$\tabularnewline
 & $-0.413\pm0.155$ & $-0.544\pm0.204$ & $-0.530\pm0.199$\tabularnewline
\hline 
\end{tabular}
\end{table}

\subsection{Comments on the results for A586}

We present our results in table \ref{tab:Virial-ratio-Master}. One
sees that the analysis of this cluster with a top hat profile is in
agreement with the findings of Ref. \cite{Bertolami:2007zm}. We observe
consistently that using weak lensing velocity dispersion (as performed
in Ref. \cite{Bertolami:2007zm}) yields a higher virial ratio. We
stress that the weak lensing velocity dispersion is not the most adequate
data source as it introduces correlations between mass and velocity
estimation. It is crucial to avoid these correlations as the aim of
this work is to detect deviations to the virial equilibrium and interpret
these as an effect due to DE-DM interaction.

\section{\label{sec:Analysis-of-the}Analysis of the A1689 cluster}

In table \ref{tab:Summary-of-A1689} we gather the data available
for A1689.
\begin{table}
\caption{\label{tab:Summary-of-A1689}Summary of A1689 available data}

\begin{tabular}{lll}
\hline 
Parameter  & Value  & Reference\tabularnewline
\hline 
$\rho_{0}$ ($10^{-25}h^{2}gr/cm^{3}$)  & $9.6\pm1.8$  & \cite{Lemze:2007gh}\tabularnewline
$r_{0}(h^{-1}kpc)$  & $175\pm18$  & \cite{Lemze:2007gh}\tabularnewline
$kT_{x}(KeV)$  & $9.2_{-0.3}^{+0.4}$  & \cite{Hoekstra:2007nc} \tabularnewline
$\sigma_{dyn}(Km/s)$  & $1172_{-99}^{+123}$  & \cite{Girardi:2000zc}\tabularnewline
$R_{dyn}(h^{-1}Mpc)$  & $2.26$  & \cite{Girardi:2000zc}\tabularnewline
\hline 
\end{tabular}

\end{table}
 As already mentioned, for this cluster we find the density and shape
parameters for the different profiles by directly fitting them to
the 2D surface density obtained from weak and strong lensing. The
relation between the 2D surface density $\kappa(R)$ and the total
3 dimensional mass density $\rho(r)$ is given by

\begin{equation}
\kappa(R)=\frac{2}{\Sigma_{crit}}\int_{R}^{\infty}\frac{\rho(r)rdr}{\sqrt{r^{2}-R^{2}}},\,
\end{equation}
where $\Sigma_{crit}$ is the critical density for lensing, which
depends on the background cosmology and also on the lens and source
sample. For the data we are considering, $\Sigma_{crit}=1.0122\, h\, g/cm^{2}$
\cite{Lemze:2007gh}, for $H_{0}=100h\, km/s/Mpc$ and observationally
$h\simeq0.7$. Then, by letting $\rho(r)$ be NFW, isothermal sphere,
Moore, or Einasto, we estimate the parameters of these four density
profiles.
\begin{table}
\caption{\label{tab:NFW-A1689} NFW fit to A1689 lensing data.}
\begin{tabular}{lll}
\hline 
Parameter  & Value  & Error\tabularnewline
\hline 
$\rho_{0}$ ($10^{-25}h^{2}gr/cm^{3}$) & $4.513$ & $0.330$\tabularnewline
$r_{0}(h^{-1}kpc)$ & $271.6$ & $12.5$\tabularnewline
$\mathcal{R}^{2}$ & $0.999$ & $-$\tabularnewline
\hline 
\end{tabular}

\end{table}

In the ensuing analysis, we will consider two methods for estimating
velocity dispersion: galaxy dynamics and X-ray temperature, as shown
in table \ref{tab:Summary-of-A1689}. While the velocity dispersion
from the galaxy dynamics is a direct measurement, converting the X-ray
temperature into a velocity dispersion involves assumptions that must
be explained. Defining $\beta=\sigma_{v}^{2}(kT/\mu m_{p})^{-1}$,
where $\mu$ and $m_{p}$ are the reduced nuclear mass and the proton
mass, allows for computing $\sigma_{v}$ knowing $kT$ once $\beta$
is specified. To first approximation one can assume the hypothesis
of density energy equipartition, which corresponds to setting $\beta=1$.
For this choice we find $\sigma_{v}=(1232\pm27)\, km/s$. Other values
for $\beta$ might be used, in particular see e.g. Ref. \cite{Girardi:2000zc}.
Another issue related to the velocity dispersion is the region in
which it is measured. In the case of $\sigma_{v}$ coming from galaxy
dynamics, the measurements are taken in a spherical region whose radius
is the distance to the centre of the outermost galaxy. For the current
cluster, this is found to be $2.26h^{-1}Mpc$. In the case of X-ray
temperature measurements, these are taken up to a radius $r_{2500}$,
i.e. the radius at which the local density is 2500 times the critical
density around the redshift of the cluster. Assuming the $\Lambda CDM$
cosmology, we find this to be of order $300h^{-1}kpc$. Notice that
the exact value of $r_{2500}$ depends on the density profile as it
is assumed, nonetheless there is an order of magnitude difference
relative to $R_{dyn}$.

\subsection{NFW profile}

In Ref. \cite{Lemze:2007gh} a joint X-ray and lensing analysis is
performed and the data has been fitted to an NFW profile, the results
of which are displayed in table \ref{tab:Summary-of-A1689}. Nonetheless
we perform our own fit to the lensing data, the results of which are
shown in table \ref{tab:NFW-A1689}.

Since the method used for computing $\rho_{K}/\rho_{W}$ for A1689
is slightly different from the one used for A586, the value and error
of the ratio will now be given in terms of the fitted parameters $\rho_{0}$
and $r_{0}$ instead of $M$ and $\left\langle R\right\rangle $.
The kinetic to potential energy density ratio is now given by%\foreignlanguage{british}{
\begin{align}
\frac{\rho_{K}}{\rho_{W}} & =-\frac{3\sigma_{v}^{2}}{8\pi G\rho_{0}r_{0}^{2}}
\frac{\left[\left(1+\frac{R}{r_{0}}\right)\ln\left(1+\frac{R}{r_{0}}\right)-\frac{R}{r_{0}}\right]\left(1+\frac{R}{r_{0}}\right)}{\left[\left(1+\frac{R}{r_{0}}\right)\left\{ \frac{1}{2}\left(1+\frac{R}{r_{0}}\right)-\ln\left(1+\frac{R}{r_{0}}\right)\right\} -\frac{1}{2}\right]},
\end{align}
%}
and the error is found by applying Eq. (\ref{eq:errRatio}) to the
previous equation. 
\begin{table}

\caption{\label{tab:Moore-A1689} Moore fit to A1689 lensing data.}%\begin{centering}
\begin{tabular}{lll}
\hline 
Parameter  & Value  & Error\tabularnewline
\hline 
$\rho_{0}$ ($10^{-25}h^{2}gr/cm^{3}$)  & $1.064$  & 0.525\tabularnewline
$r_{0}(h^{-1}kpc)$  & $357$  & 102\tabularnewline
$\mathcal{R}^{2}$  & $0.973$ & -\tabularnewline
\hline 
\end{tabular}
%\par\end{centering}

\end{table}
\begin{table}

\caption{\label{tab:Einasto-A1689} Einasto fit to A1689 lensing data.}%\begin{centering}
\begin{tabular}{lll|ll}
 & \multicolumn{2}{c}{$n=1$} & \multicolumn{2}{|c}{$n=6$}\tabularnewline
\hline 
Parameter  & Value  & Error & Value  & Error\tabularnewline
\hline 
$\rho_{0}$ ($10^{-25}h^{2}gr/cm^{3}$)  & $46.49$  & $5.165$  & $66005$ & $7358$\tabularnewline
$r_{0}(h^{-1}kpc)$  & $122.5$  & $8.47$  & $467.3$ & $31.99$\tabularnewline
$\mathcal{R}^{2}$  & $0.9834$  & - & 0.998 & -\tabularnewline
\hline 
\end{tabular}
%\par\end{centering}

\end{table}

For completeness we compute the ratio $\rho_{K}/\rho_{W}$ using our
best fit and the fit found in Ref. \cite{Lemze:2007gh}. The results
are displayed in table \ref{tab:Virial-ratio-Master2}.

\subsection{Isothermal sphere density profile}

\begin{figure}[t]
\includegraphics[width=0.45\textwidth]{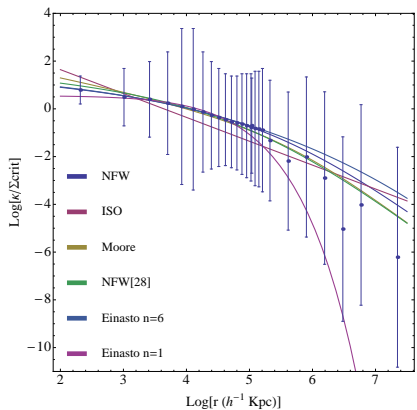}

\caption{\label{fig:kappa}Fits to 2D surface density for A1689.}
\end{figure}

Following the procedure described above, we fit the projected mass
data obtained from lensing measurements to the isothermal mass profile
given by Eq. (\ref{eq:Iden}). One should notice that the isothermal
profile, unlike the NFW, Moore and Einasto profiles, has only one
free parameter, namely $\rho_{0}r_{0}^{2}$. The fit to the data yields
$\rho_{0}r_{0}^{2}=(3.987\pm0.333)\times10^{-21}$ with $\mathcal{R}^{2}=0.845$.
As indicated by the low value of $\mathcal{R}^{2}$ the isothermal
profile is not very suitable to describe the mass distribution in
the cluster A1689. This can also be seen in Figure \ref{fig:kappa}.
The energy density ratio is given in terms of the fitted parameter
and the velocity dispersion by: 
\begin{equation}
\frac{\rho_{K}}{\rho_{W}}=-\frac{3}{8\pi}\frac{\sigma_{v}^{2}}{Gr_{0}^{2}\rho_{0}},\label{eq:ratioISO}
\end{equation}
 with an error given by Eq. (\ref{eq:errRatio}). Note that Eq. (\ref{eq:ratioISO})
is independent of the region where the velocity dispersion is measured,
$R$. This is a characteristic feature of this profile. In Table \ref{tab:Virial-ratio-Master2}
we display the results found by applying the isothermal profile to
the study of A1689.

\subsection{Moore's profile}

Repeating the above procedure for the Moore's profile with data depicted
in Table \ref{tab:Moore-A1689} we get the results shown in Table
\ref{tab:Virial-ratio-Master2} from which we can see that no signal
of DE-DM interaction can be seen as no deviation from the virial ratio
value $-1/2$ is unambiguously seen.

\subsection{Einasto's profile}

Repeating the above procedure for the Einasto profile with data depicted
in Table \ref{tab:Einasto-A1689} we obtain the results shown in Table
\ref{tab:Virial-ratio-Master2}.

\begin{table}
\caption{\label{tab:Virial-ratio-Master2}Virial ratio from the various observations
of A1689 given by Refs. \cite{Lemze:2007gh,Hoekstra:2007nc,Girardi:2000zc}.}
\begin{tabular}{l|l|l|l|l}
\hline 
Method  & NFW  & NFW \cite{Lemze:2007gh}  & Isothermal  & \tabularnewline
\hline 
X-ray Temperature  & $-0.518\pm0.060$ & $-0.538\pm0.144$ & $-0.715\pm0.067$ & \tabularnewline
Velocity distribution  & $-0.408\pm0.099$ & $-0.485\pm0.173$ & $-0.647\pm0.146$ & \tabularnewline
\hline 
\end{tabular}

\noindent %
\begin{tabular}{l|l|l|l}
\hline 
 & Moore  & Einasto $n=1$ & Einasto $n=6$\tabularnewline
\hline 
 & $-0.549\pm0.391$  & $-0.527\pm0.093$ & $-0.520\pm0.086$\tabularnewline
 & $-0.479\pm0.389$  & $-0.473\pm0.130$ & $-0.353\pm0.098$\tabularnewline
\hline 
\end{tabular}
\end{table}

\section{\label{sec:Interaction-and-violation}DE-DM Interaction}

In this section we relate the virial ratios computed above to the
interaction parameter between DM and DE following Ref.~\cite{Bertolami:2007zm}.
We start by briefly reviewing how the interaction is parametrised
and how this parameter is related to the virial ratio in relaxed structures.
In Ref.~\cite{Bertolami:2007zm} the interaction was analysed within
the context of two distinct models of interacting DE: coupled quintessence
\cite{Amendola:1999er} and the generalized Chaplygin Gas (GCG) \cite{Bento:2002ps}.
It was shown that for $\omega_{DE}=-1$, the coupling parameter of
the interacting quintessence, $\xi$ (cf. below), is related to the
$\alpha$ parameter of the GCG equation of state $p=-A/\rho^{\alpha}$,
by the scaling parameter (defined below) $\eta=3(1+\alpha)$. Keeping
this mapping between the two distinct models in mind, we review how
to relate virial ratio the interaction parameter in the context of
the coupled quintessence model.\\
 The conservation equations for DM and DE read 
\begin{equation}
\dot{\rho}_{DM}+3H\rho_{DM}=\xi H\rho_{DM},\label{eq:DMCons}
\end{equation}
 
\begin{equation}
\dot{\rho}_{DE}+3H\rho_{DE}(1+\omega_{DE})=-\xi H\rho_{DM}.\label{eq:DECons}
\end{equation}

It is assumed that there is a scaling behavior between the DE and
DM energy densities, 
\begin{equation}
\frac{\rho_{DE}}{\rho_{DM}}=\frac{\Omega_{DE}}{\Omega_{DM}}a^{\eta},
\end{equation}
 where $\eta$ describes the scaling and is related to the coupling
$\xi$ by \cite{Bertolami:2007zm} 
\begin{equation}
\xi=-\frac{\eta+3\omega_{DE}}{1+(\Omega_{DM0}/\Omega_{DE0})a^{-\eta}}.
\end{equation}

Following the derivation of the Layzer-Irvine equation in Ref.~\cite{Peebles:1993},
noting that the scaling of the matter kinetic energy with the scale
factor is unchanged by the presence of interaction with DE, $\rho_{K}\propto a^{-2}$,
and that the interaction will modify the scaling of the potential
energy with $a$ to $\rho_{W}\propto a^{\xi-1}$ one finds \cite{Bertolami:2007zm}
\begin{equation}
\dot{\rho}_{DM}+H(2\rho_{K}+\rho_{W})=\xi\rho_{W}H.
\end{equation}
 This is the modified Layzer-Irvine equation. Assuming that locally
$\dot{\rho}\approx0$, we find that the virial theorem is modified
to 
\begin{equation}
\frac{\rho_{K}}{\rho_{W}}=-\frac{1}{2}+\frac{\xi}{2},
\end{equation}
 where $\xi$ is the interaction strength as defined in Eqs. (\ref{eq:DMCons})
and (\ref{eq:DECons}).

We are therefore mapping deviations from virial equilibrium to the
existence of DE-DM interaction. If the data on a particular cluster
yields $\rho_{K}/\rho_{W}>-1/2\Rightarrow\xi>0$ and the energy transfer
occurs from DE to DM. Conversely $\rho_{K}/\rho_{W}<-1/2\Rightarrow\xi<0$
and the energy transfer occurs from DM to DE.

As already stressed in the previous section, the results in tables
\ref{tab:Virial-ratio-Master} and \ref{tab:Virial-ratio-Master2}
do not point to a preferred direction of the deviation of the ratio
from its canonical value of $1/2$. Therefore there is no preferred
sign for $\xi$ and the results are largely consistent with $\xi=0$.
This contrasts with the results found in \cite{Bertolami:2007zm,Bertolami:2007tq}
where the analysis of the A586 cluster with a top hat profile and
weak lensing inferred velocity dispersion pointed towards $\xi<0$.
In fact, this suggests that changing the density profile and considering
the related methodological difficulties do not allow for a definite
conclusion on the virial ratio. We believe that the current analysis
represents a step forward relative to the original proposal as it
deals with more realistic density profiles and uses uncorrelated data
sources in the estimation of the velocity dispersion. However, in
order to distinguish between the different physical scenarios $\xi>0$,
$\xi<0$ and $\xi=0$ it is crucial to have a better understanding
both of the mass distribution within the clusters and a more importantly
a more accurate knowledge of their dynamical state. This will allow
to decrease the error bars and detect putative deviations of the ratio
from $-0.5$.

\section{\label{sec:Conclusions}Conclusions}

In this work we have analyzed the virial equilibrium of clusters A586
and A1689 using four density profiles, namely NFW, isothermal spheres
and Moore's and Einasto's profiles. The method employed represents
an evolution from the original proposal as it deals with more realistic
mass distributions. For the A586 cluster we have only found evidence
of deviation of the virial ratio $\rho_{K}/\rho_{W}=-1/2$ for the
over-simplistic Top-Hat density profile, in accordance with \cite{Bertolami:2007zm}.
In what concerns cluster A1689, we have not encountered a convincing
evidence of this interaction except when using the uni-parametric
isothermal sphere distribution. This could reflect the claims on A1689
that it is not a relaxed cluster, in particular, that it exhibits
signs of triaxiality \cite{Sereno:2011qk}. Despite the fact that
search for interaction with more realistic mass profiles has not returned
a clear signal in either cluster, one should stress that the encountered
error bars are still too large at present. In order to study the interaction
between DE and DM via deviations from the virial equilibrium it is
crucial to have a better understanding of the cluster mass distribution
and of their dynamical state. We believe N-body simulations have a
crucial role to play in this problem.

\subsection*{Acknowledgments}

\vspace{0.3cm}

\noindent The authors would like to thank Felipe Andrade Santos for
discussions on the A1689 X-ray substructure, and Doron Lemze, Keiichi
Umetsu and Adi Zitrin for several discussions and for kindly allowing
us to use their data of cluster A1689. The work of O.B. is partially
supported by the Funda\c{c}\~{a}o para a Ci\^{e}ncia e a Tecnologia
(FCT) project PTDC/FIS/111362/2009. The work of F.G.P. is supported
by the FCT grant SFRH/BD/35756/2007. The work of M.L.D. is supported
by CSIC (Spain) under the contract JAEDoc072, with partial support
from CICYT project FPA2006-05807, at the IFT, Universidad Autonoma
de Madrid, Spain.

%\cite{Bernardini:2008pn}

\end{document}